\documentclass[9pt,conference]{IEEEtran}
\IEEEoverridecommandlockouts
\usepackage{algpseudocode}
\usepackage{cite}
\usepackage{graphicx}
\usepackage{circledsteps}
\usepackage{color}
\usepackage{enumitem}
\usepackage{multirow}
\usepackage{etoolbox,siunitx}
\usepackage{booktabs}
\usepackage{balance}
\usepackage{amssymb}
\usepackage{bm}
\usepackage{arydshln}
\usepackage{caption}
\usepackage{amsmath,amssymb,amsfonts}
\usepackage{graphicx}
\usepackage{textcomp}
\usepackage{url}

\providecommand{\yz}[1]{\textcolor{black}{{#1}}}

\begin{document}	
\title{Piano Transcription by Hierarchical Language Modeling with Pretrained Roll-based Encoders}

% \author{\IEEEauthorblockN{Di Li\IEEEauthorrefmark{1}, Yongyi Zang\IEEEauthorrefmark{1} and Qiuqiang Kong\IEEEauthorrefmark{1}}
% 	   \IEEEauthorblockA{\IEEEauthorrefmark{1}The Chinese University of Hong Kong, Hong Kong SAR, China\\
% 		Email: \{emails}}

\author{\IEEEauthorblockN{Dichucheng Li}
\IEEEauthorblockA{\textit{DSP Lab, Electrical Engineering Dept.} \\
\textit{The Chinese University of Hong Kong}\\
Hong Kong SAR, China \\
dccli21@m.fudan.edu.cn}
\and
\IEEEauthorblockN{Yongyi Zang}
\IEEEauthorblockA{\textit{Independent Researcher} \\
Seattle, WA, USA \\
zyy0116@gmail.com}
\and
\IEEEauthorblockN{Qiuqiang Kong$^{\dagger}$\thanks{$\dagger$ Corresponding author}}
\IEEEauthorblockA{\textit{DSP Lab, Electrical Engineering Dept.} \\
\textit{The Chinese University of Hong Kong} \\
Hong Kong SAR, China \\
qqkong@ee.cuhk.edu.hk}
}

\maketitle
\begin{abstract}
% Automatic Music Transcription (AMT), aiming to get musical notes from raw audio, has traditionally employed two distinct approaches: frame-level systems with piano-roll outputs and language model (LM)-based systems with direct note-level predictions. However, the former method requires manual thresholding for note event detection and the latter one struggles with long sequence processing. In this paper, we propose a novel hybrid method that integrates pre-trained roll-based encoders with an LM decoder to leverage the strengths of both methods. Besides, our approach implements a hierarchical prediction strategy: first predicting onset and pitch, followed by velocity, and finally offset. It reduces computational costs by decomposing a long token sequence into short sequences of different hierarchies. We evaluate our methods using two benchmark roll-based encoders, where our approach outperforms the piano-roll outputs of both encoders. This demonstrates that our methods can serve as a plug-in to enhance the performance of any roll-based encoder. Empirical analysis reveals that encoder choice significantly outweighs language model size in impact on performance, with larger language models showing diminishing returns. 

Automatic Music Transcription (AMT), aiming to get musical notes from raw audio, typically uses frame-level systems with piano-roll outputs or language model (LM)-based systems with note-level predictions. However, frame-level systems require manual thresholding, while the LM-based systems struggle with long sequences. In this paper, we propose a hybrid method combining pre-trained roll-based encoders with an LM decoder to leverage the strengths of both methods. Besides, our approach employs a hierarchical prediction strategy, first predicting onset and pitch, then velocity, and finally offset. The hierarchical prediction strategy reduces computational costs by breaking down long sequences into different hierarchies. 
Evaluated on two benchmark roll-based encoders, our method outperforms traditional piano-roll outputs 0.01 and 0.022 in onset-offset-velocity F1 score, demonstrating its potential as a performance-enhancing plug-in for arbitrary roll-based music transcription encoder. 
% We release the code of this work at \url{https://github.com/yongyizang/AMT_train}
% Evaluated on two benchmark roll-based encoders, our method outperforms traditional piano-roll outputs, demonstrating its potential as a performance-enhancing plug-in for any roll-based encoder. 

% Empirical analysis highlights that encoder choice significantly impacts performance more than LM size, with larger models offering diminishing returns.
% \yz{Automatic Music Transcription (AMT) has traditionally employed two distinct approaches: frame-level systems with piano-roll outputs, which require manual thresholding for note event detection, and language model (LM)-based systems with direct note-level predictions, which struggle with long sequence processing. We propose a novel hybrid method that bridges this divide, integrating pre-trained roll-based encoders with an LM decoder. Our approach implements a hierarchical prediction strategy for note events, enhancing efficiency by reducing sequence length, showing improved performance. Empirical analysis reveals that encoder choice significantly outweighs language model size in impact on performance, with larger language models showing diminishing returns. 
% % This study highlights the crucial role of encoder selection in AMT and calls for further investigation into improving the scalability of LM-based AMT systems.
% }
\end{abstract}

\vspace{-0.2cm}
\section{Introduction}
\label{sec:intro}
% It is a challenging task due to the high polyphony of piano music. 

Automatic music transcription (AMT) is a task of converting audio recordings into symbolic representations~\cite{benetos2018automatic}. As a key topic in music information retrieval (MIR), AMT bridges audio-based and symbolic-based music understanding. AMT systems enable applications such as score following~\cite{li2016approach} and audio-score alignment~\cite{niedermayer2010multi}. 

Piano transcription, an instrument-specific subtask of AMT, is a challenging task due to the high polyphony of piano music. Numerous methods have been utilized for piano transcription in recent decades, including Factorization-based models~\cite{khlif2015iterative}, adaptive estimation of harmonic spectra~\cite{vincent2009adaptive}, and SVM-HMM structure~\cite{nam2011classification}. 
With deep learning's rise, models like CNN~\cite{kelz2016potential} and CRNN~\cite{sigtia2016end} have been effectively applied in AMT. Onsets \& Frames system~\cite{onsets} significantly improved note-level metrics by integrating onset and pitch detection. Kong et al.~\cite{kong2021high} further enhance the AMT system by proposing a high-resolution AMT system trained by regressing precise onset and offset times of piano notes. To minimize model size, studies~\cite{wei2022hppnet, fernandez2023triad} have used prior knowledge of harmonic structures in audio representations to develop Dilated Convolutional networks. 
Transformer is a revolutionary model with a encoder-decoder architecture, the self-attention mechanism of which can extract global features and catch long-term relationships. 
% Wu et al.~\cite{wu2024piano} proposed a harmonic mask self-attention for better capturing harmonic features.
Toyama et al.~\cite{toyama2023automatic} proposed a two-level hierarchical frequency-time Transformer to catch long-term spectral and temporal dependencies to determine the precise onset and offset for each note. The above methods use piano rolls as output, which has a frame-level resolution and requires a threshold and a post-processing stage to decode it into a note sequence.
Hawthorne et al.~\cite{hawthorne2021sequence} revolutionize the AMT framework by treating it as a sequence-to-sequence task, where the the output is directly a sequence of note-event tokens, eliminating the need for extensive threshold-based post-processing

\yz{Frame-level and language model (LM)-based systems have traditionally been viewed as distinct approaches in AMT. Frame-level systems utilize a compact piano-roll objective but require complex post-processing, while LM-based systems directly output note-level predictions. However, LM-based systems face challenges due to the lengthy sequences created by flattening note events that contain tokens of onset, offset, pitch, and velocity, resulting in resource-intensive training and inference processes. Furthermore, audio encoder selection has been explored in related domains such as audio captioning~\cite{mei2021audio, mei2024wavcaps, zhu2024cacophony, kim2024enclap, xu2024efficient} and multimodal large language models~\cite{chen2024internvl, chu2023qwen, liu2024music, deng2023musilingo}, yet its impact on AMT remains unexplored. This gap in research presents an opportunity to bridge the divide between frame-level and LM-based approaches.}
%, potentially leading to more efficient and effective AMT systems.

% In this paper, we proposed a method that combines the advantages of frame-based models and seq-to-seq methods for piano transcription.  The input to the model is a sequence of audio embedding frames extracted by a pretrained AMT encoder followed by a task token and a sequence of ground-truth tokens representing all notes in this piece of audio. We train the decoder and finetune the encoder in a hierarchical way using the query token to control the task order: we firstly predict the onset and pitch of all notes in the input audio simultaneously, then the onset and pitch prediction token sequence of each note are utilized as the input of the decoder to predict velocity and offset of each note respectively. Our model outperforms all of other models. It is worth noting that our models outperform the performance of directly outputting piano roll predictions using their respective encoder. Thus, any trained encoder can be plugged into our decoder structure to enhance its transcription performance.

\yz{In this paper, we introduce a novel approach that leverages the strengths of both roll-based systems and note-based language models for music representation. Our proposed method employs pre-trained roll-based systems as encoders and a language model as a decoder, effectively combining their respective advantages. To enhance prediction efficiency, we implement a hierarchical prediction strategy: first predicting onset and pitch, followed by velocity, and finally offset. To achieve the hierarchical prediction strategy, we trained three models with the same model architecture to predict onset and pitch, velocity, and offset, respectively. This approach significantly reduces the prediction sequence length compared to a flattened note event sequence, resulting in improved performance.} \yz{We empirically evaluate the impact of different roll-based encoders and language model decoder size. Our findings reveal that the choice of encoder has a much more substantial effect on overall performance than the size of the language model. Notably, we observe that increasing the language model size does not reflect as improved model performance, corroborating observations reported in~\cite{hawthorne2021sequence}. We further found that velocity modeling is more prone to overfitting compared to onset-pitch and offset. Our findings highlight the importance of encoder selection in AMT tasks, calling for further research in improving the scalability of language-model-based AMT systems.  }

\yz{This paper is structured as follows: Section \ref{sec:amt_sys} provides a detailed description of both the piano roll-based and LM-based systems. In Section \ref{sec:flattened}, we compare the traditional flattened token construction method with our proposed hierarchical approach. Our experimental design is outlined in Section \ref{sec:exps}, followed by Section \ref{sec:results}, which presents our findings, including results, ablation studies, and related discussions. Finally, Section \ref{sec:conclusions} offers concluding remarks. }

\section{Roll-based and LM-based AMT Systems}
\label{sec:amt_sys}
\begin{figure}
    \centering
    \includegraphics[width=0.95\linewidth]{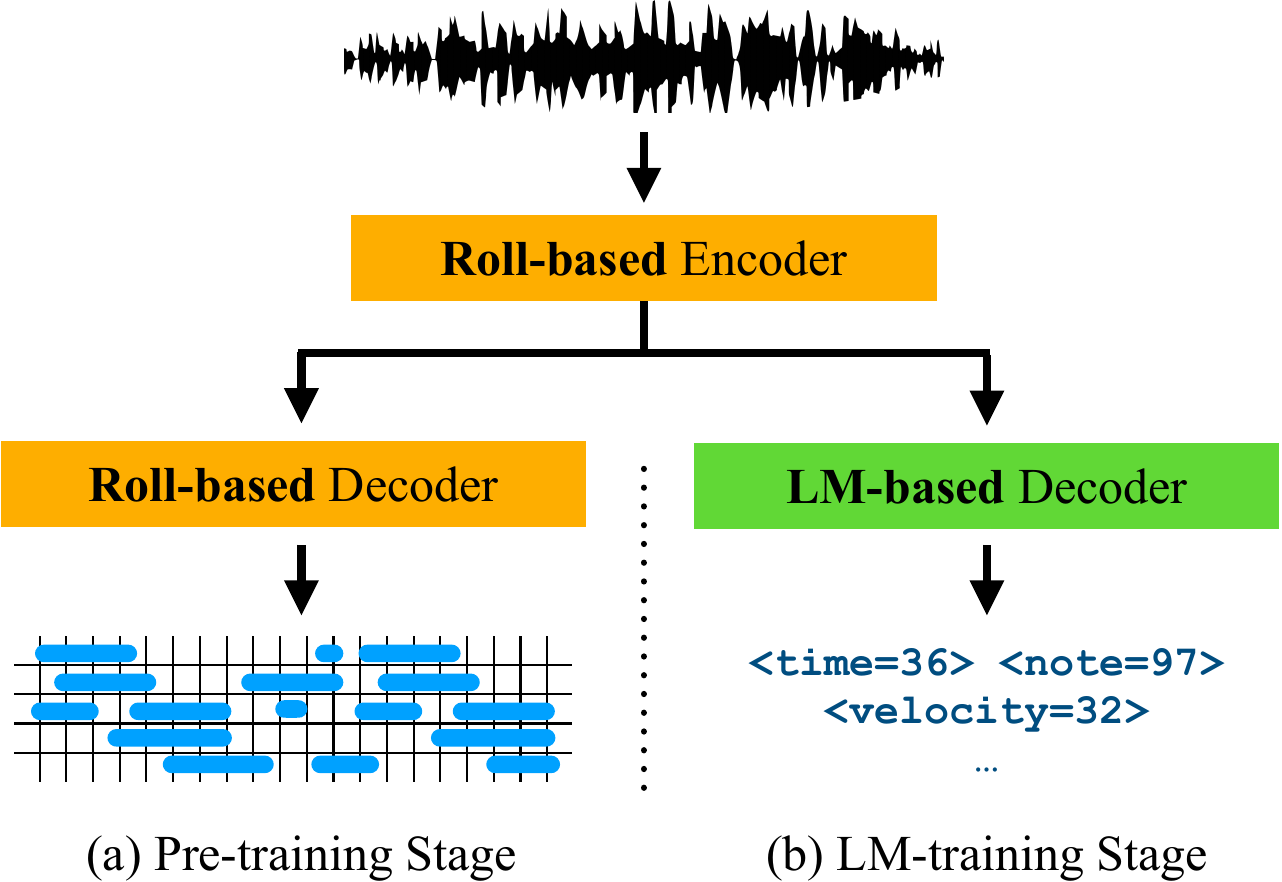}
    \caption{Proposed system architecture. (a) Pre-training stage trains the roll-based encoder with frame-level objectives. (b) LM-training stage connects it to a LM decoder, training with note-level objectives.}
    \label{fig:system-arch}
\end{figure}

\subsection{Roll-based AMT Systems}
\yz{In roll-based AMT, a waveform $x$ is firstly transformed to an feature in time-frequency domain $X \in \mathbb{R}^{T \times F}$, where $T$ is time frames and $F$ is frequency bins, using short-time Fourier transform (STFT). Then, the feature is transformed to predict a piano roll $Y \in \{0,1\}^{T \times K}$ , where $K$ is 88 possible pitches, and 0 or 1 encodes absence or presence of each pitch. The neural network model $f_\theta(X)$ predicts $\hat{Y} \in [0,1]^{T \times K}$, representing predicted pitch probabilities. Training typically uses binary cross-entropy loss:}
\begin{equation}
\mathcal{L}_\text{BCE} = -\frac{1}{TK} \sum_{t=1}^T \sum_{k=1}^K [Y_{t,k} \log(\hat{Y}_{t,k}) + (1-Y_{t,k}) \log(1-\hat{Y}_{t,k})].
\end{equation}
\yz{At inference time, the continuous predictions $\hat{Y}_{t,k}$ are typically thresholded to obtain binary predictions, which are then post-processed to extract note events with onset and offset times. While roll-based systems have shown good performance, the need for post-processing limits their use cases, and have motivated the development of LM-based systems.}

\subsection{LM-based AMT Systems}

\yz{Language Model (LM)-based AMT systems treat music transcription as a sequence generation task. In these systems, the input audio $X \in \mathbb{R}^{T \times F}$ is typically first encoded into a sequence of hidden representations $H \in \mathbb{R}^{T' \times D}$, where $T'$ is the number of encoded time steps and $D$ is the dimension of the hidden representation. The system then generates a sequence of note events $Y = (y_1, ..., y_N)$, where each $y_i$ represents a note event typically consisting of onset time, pitch, duration, and velocity. The neural network model in an LM-based system can be represented as a conditional language model $p_\theta(Y|X)$, where $\theta$ are the learnable parameters. This model generates the probability distribution of the next note event given the previous events and the input audio:}
\begin{equation}
p_\theta(Y|X) = \prod_{i=1}^N p_\theta(y_i|y_{<i}, X).
\end{equation}
\yz{Therefore, the training objective for LM-based systems typically uses the negative log-likelihood loss:}
\begin{equation}
\mathcal{L}_\text{NLL} = -\frac{1}{N} \sum_{i=1}^N \log p_\theta(y_i|y_{<i}, X).
\end{equation}
\yz{At inference time, the model generates note events autoregressively, often using beam search or other decoding strategies to improve the quality of the generated sequence. This approach allows for direct generation of note-level predictions without the need for post-processing, potentially capturing long-term dependencies in the music. While LM-based systems offer the advantage of direct note-level prediction, they are often computationally expensive due to the need for flattening of note event sequence.}

\section{Flattened and Hierarchical Token Structure}
\label{sec:flattened}
\begin{figure}
    \centering
    \includegraphics[width=0.95\linewidth]{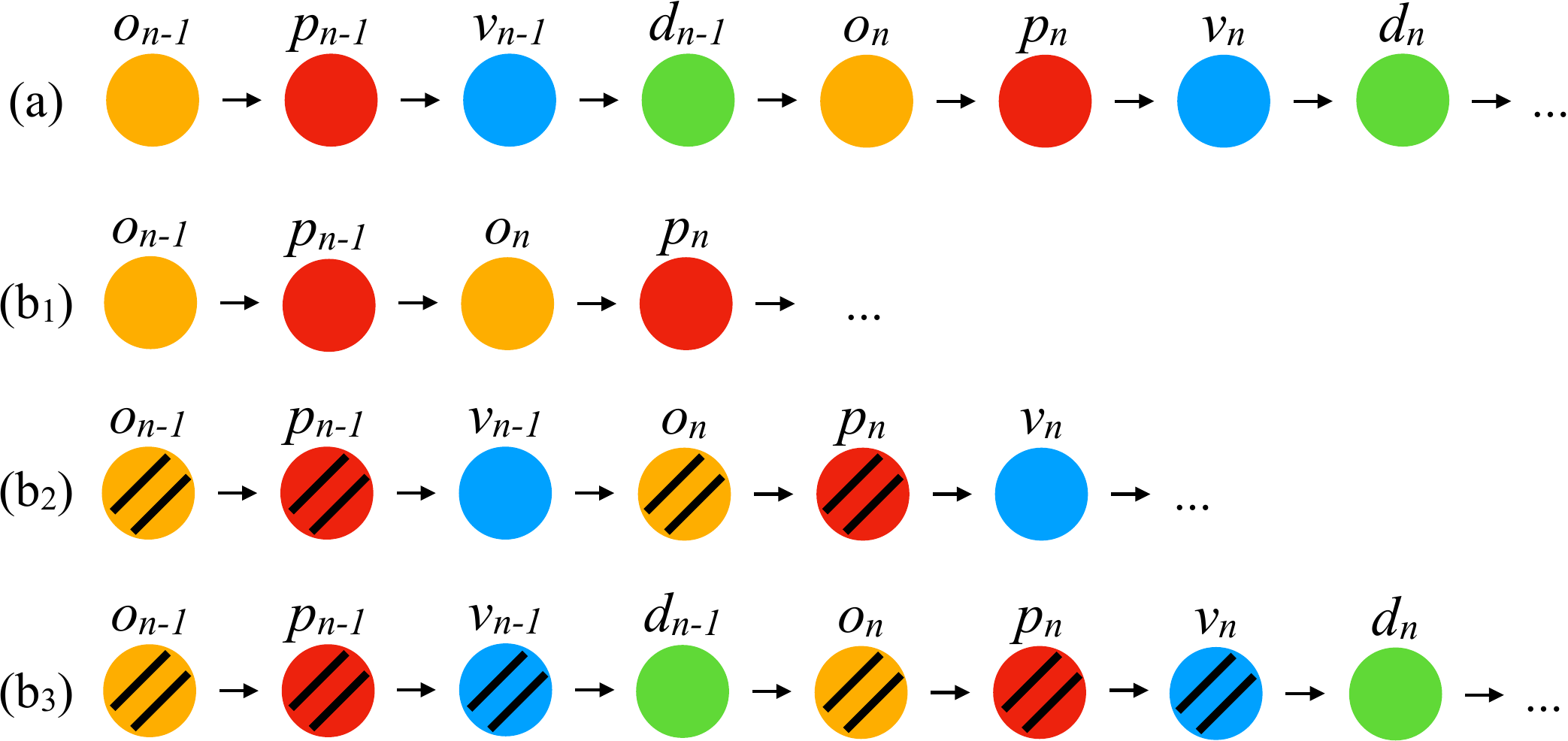}
    \caption{Flattened and hierarchical token sequence. $o_n, p_n, v_n$ and $d_n$ represents a note event with onset time $o_n$, pitch $p_n$, velocity $v_n$, and offset $d_n$. Flattened approach (a) forms one sequence, while the hierarchical approach ($b_1$, $b_2$ and $b_3$) forms three sequences. Two lines on $o$ and $p$ tokens denote that they are fixed from $b_1$ during $b_2$ and $b_3$. Best viewed in color. See Sec.~\ref{sec:flattened} for details.}
    \label{fig:token-structure}
\end{figure}
\yz{Our system is composed of an encoder $f_\text{enc}$ and a LM-based decoder $f_\text{dec}$, as illustrated in Figure~\ref{fig:system-arch}. The encoders $f_\text{enc}$ are pre-trained on piano roll objectives, and transforms the input audio $X \in \mathbb{R}^{T \times F}$ into a sequence of hidden representations $H = f_\text{enc}(X) \in \mathbb{R}^{T' \times D}$, where the LM decoder $f_\text{dec}$ is designed to predict note-level output $Y = (y_1, ..., y_N)$ from the frame-level feature $H$ extracted by the encoder. We propose a hierarchical prediction strategy for note events, where $f_\text{dec}$ predicts note pitch, velocity, and offset sequentially, controlled by task-specific query tokens $q \in \{q_p, q_v, q_f\}$ with a vocabulary size of three. The probability of generating a note event $y_i$ can be expressed as:}
\begin{equation}
p_\theta(y_i|y_{<i}, X, q) = f_\text{dec}(y_{<i}, H, q),
\end{equation}
\yz{where $\theta$ represents the learnable parameters of both the encoder and decoder. In this section, we detail how we construct the token sequence used to train our system.}

% The input to our LM decoder is a sequence of 10-second audio embedding frames extracted by a pretrained encoder followed by a query token sequence and a sequence of target tokens representing all notes in the input piano music. Moreover, we need a sequence of mask tokens to let the decoder know which part to predict.

% \begin{table}
%   \caption{Token Dictionary}
%   \vspace{6pt}
%   \label{tab:dic}
%   \centering
%   \begin{tabular}{lccc}
%     \toprule
%     Token Class & Token Class Size \\
%     \midrule
%     Special Token & 4 \\
%     Task & 3 \\
%     Time & 1001 \\
%     Pitch & 128 \\
%     Velocity & 128 \\
%     Note Sustain & 1 \\
% \bottomrule
% \end{tabular}
% \end{table}

% \subsection{Token Dictionary}
% As shown in \yz{Table} \ref{tab:dic}, 
Let $H = (x_{1:T'})$ be the hidden representations extracted from the input audio sequence, and $Y = y_{1:N}$ be the sequence of note events, where each $y_n = (o_n, p_n, v_n, d_n)$ represents a note event with onset time $o_n$, pitch $p_n$, velocity $v_n$, and offset $f_n$. For $p_{\text{onset-pitch}}$, $p_{\text{velocity}}$, and $d_{\text{offset}}$, we add \texttt{<sos>} to the sequence and append task-specific query tokens $q_{p}$, $q_v$, or $q_d$. Notes are organized by onset time (first to last), then pitch (low to high). We use \texttt{<eos>} to end sequences and \texttt{<pad>} for batching. For flattened sequences, we follow~\cite{hawthorne2021sequence}, organizing note onset, pitch, velocity, and offset similarly. 

% We employ LLaMA~\cite{touvron2023LLaMA} as our language model architecture, which is a decoder-only model

Differently than~\cite{hawthorne2021sequence}, which employs an encoder-decoder language model T5~\cite{raffel2020exploring}, we use LLaMA~\cite{touvron2023llama}, a decoder-only language model architecture. The primary difference between these architectures lies in how they process the input sequence to generate the output sequence. The encoder-decoder architecture processes the previously generated tokens $y_{1:{n-1}}$ using an encoder into a sequence of hidden representations $H_{\text{enc}}$, which is used to predict the next token $y_n$, i.e. $p(y_n) = p(y_n | H_{\text{enc}})$. In contrast, a decoder-only architecture generates the output sequence by directly conditioning on the previous tokens without needing an explicit encoder step. The model autoregressively predicts the next token from the preceding tokens in the sequence, i.e. $p(y_n) = p(y_n | x_{1
}, y_{1:{n-1}})$.

A critical aspect of the model is its ability to encode positional information, as the Transformer architecture itself is agnostic to the order of tokens. To address this, we use two types of positional encodings: absolute positional encodings for audio embeddings and rotary position embeddings (RoPE)~\cite{su2024roformer} for note event tokens.

Absolute positional encodings introduce explicit position information to the input tokens by adding a fixed positional vector to each token embedding. The positional vector is calculated using sine and cosine functions of varying wavelengths, which ensure that each position in the sequence has a unique encoding. Formally, for a position $pos$ and dimension $i$, the positional encoding is defined as:
\begin{equation}
\begin{split}
\text{PE}(pos, 2i) &= \sin\left(\frac{pos}{10000^{\frac{2i}{d}}}\right), \\
\text{PE}(pos, 2i+1) &= \cos\left(\frac{pos}{10000^{\frac{2i}{d}}}\right),
\end{split}
\end{equation}
where $d$ is the dimensionality of the embeddings. This encoding is added directly to the token embeddings, allowing the model to differentiate between tokens based on their positions in the sequence.

In contrast, Rotary Position Embeddings (RoPE)~\cite{su2024roformer} offer a more flexible and efficient way of encoding positional information, especially for tasks involving long-range dependencies. RoPE modifies the query ($Q$) and key ($K$) vectors by applying a rotation in the embedding space. This introduces a relative positioning between tokens while preserving the inner-product structure of the attention mechanism. The rotary transformation is applied as follows:
\begin{equation}
\begin{split}
Q'_i &= Q_i \cos(\theta) + Q_{i+1} \sin(\theta), \\
K'_i &= K_i \cos(\theta) + K_{i+1} \sin(\theta),
\end{split}
\end{equation}
where $\theta$ is a position-dependent angle. Recall the probability of generating the entire sequence $Y$ is given by:
\begin{equation} p(Y) = \prod_{n=1}^N p(y_n | x_{1
}, y_{1:{n-1}}). \end{equation}
In our hierarchical model, we extend this with task-specific query tokens $q$. The probability of generating each note $y_n$ is now factorized into subcomponents, each handled by a separate language model:
\begin{equation} \begin{split} p(y_n) &= p(o_n, p_n | x_{1
}, y_{1:{n-1}}, q_{p})  \cr \times &p(v_n | x_{1
}, y_{1:{n-1}}, o_n, p_n, q_v)  \cr \times &p(d_n | x_{1
}, y_{1:{n-1}}, o_n, p_n, v_n, q_f). \end{split} \end{equation}
Here, $p_{\text{onset-pitch}}$, $p_{\text{velocity}}$, and $p_{\text{offset}}$ represent distinct language models for onset-pitch, velocity, and offset predictions, respectively. These models utilize the attention mechanism described earlier, where the query, key, and value matrices are computed from the input sequence and prior predictions. The use of separate language models for each aspect of the musical event sequence reduces interference between tasks and enables more focused learning. Each language model follows the same Transformer-based structure, where $W_p$ is a learnable linear projection matrix:
\begin{equation} p_{\text{model}}(\cdot) = \text{softmax}(W_p \cdot \text{FFN}(\text{MultiHead}(Q, K, V))). \end{equation}
Here, $FFN$ represents Feed Forward Network. This formulation allows each submodel to specialize in a specific attribute of the musical sequence while maintaining a consistent underlying architecture. The model benefits from this hierarchical structure, as predictions for onset-pitch, velocity, and offset are made sequentially, with each prediction conditioned on the previous ones.

The efficiency of this hierarchical approach is reflected in its time complexity. For each language model, the time complexity remains $O((T+N)^2D)$, where $T$ is the length of the input sequence, $N$ is the length of the note sequence, and $D$ is the hidden dimension. However, the overall complexity is reduced to $O(3(T+N)^2D)$ by splitting the task across three separate models. This is a significant improvement over the $O((T+3N)^2D)$ complexity that would result from using a single model for all tasks. Given that $T \ll N$, this leads to an almost threefold reduction in time complexity, greatly improving the model’s efficiency while maintaining high accuracy.

\section{Experiment Design}
\label{sec:exps}
\subsection{Encoders}
% We choose three SOTA piano transcription models (CRNN, HPPNet, HFT-transformer) as our backbone encoder respectively. 
\yz{For a comprehensive evaluation, we employ two benchmark roll-based systems, and adapt them as encoder. Specifically, we use CRNN~\cite{kong2021high} and HPPNet~\cite{wei2022hppnet}.} \yz{CRNN is comprised of convolutional layers, followed by bi-directional GRU layers and a linear readout layer. Two models of the same architecture are designed to perform note and pedal predictions. We take the embedding before the final readout layer, with 768 dimensions, and concatenate both to form a 1536-dim embedding as the representation $H$ into the LM decoder.} \yz{HPPNet combines convolutional and recurrent elements, introducing "harmonic dilated convolution" (HD-Conv) layers to exploit harmonic characteristics of the input constant-Q transform. }
%These layers extract ratios above and below a target fundamental frequency for each pitch band. The architecture includes two acoustic models with HD-Conv layers and four readout layers consisting of frequency-grouped LSTM (FG-LSTM) layers for onset, frame, offset, and velocity detection. The audio representation $H$ is constructed by flattening the 88-key representation across all readout layers, yielding a 352-dimensional embedding.}

% \subsubsection{CRNN}
% The CRNN model~\cite{kong2021high} we used is composed of convolutional layers followed by bidirectional gated recurrent units (biGRU) layers and one fully connected layer. This model predicts note onset, pitch, offset, and velocity simoutaneously. We concatenated four embeddings of onset, pitch, offset, and velocity before fully connected layers along channel dimension as the input of the decoder.
% \subsubsection{HPPNet}
% The hppnet was proposed in~\cite{wei2022hppnet}. The overall design of this system is also CRNN. It leverages the harmonic features of the spectrum and the time-invariance of piano note pitches to design Harmonic Dilated Convolution and Frequency Grouped Recurrent Neural Network, which reduce the model’s parameter while more efficiently capturing meaningful information from the input features, thereby improving the accuracy of piano transcription. We concatenated the output of the four predictions of onset, offset, velocity, and pitch as the input of the LM decoder.

\subsection{Token Dictionary}
% Our token dictionary consists of six classes of tokens that are "Special Tokens", "Time", "Pitch", "Velocity", "Task", and "Note Sustain". "Special Tokens" contains four tokens that represents "pad", "start of sentence", "end of sentence", and "unknown". "Time" contains 1001 tokens representing which frame the current note event is in (10ms per frame). "Pitch" contains 128 tokens representing the pitch of the current note. "Velocity" contains 128 tokens representing the velocity of the current note. "Task" is composed of three tokens representing "onset prediction", "offset prediction", and "velocity prediction", respectively. "Note sustain" is a token that represents whether the onset or offset of the current note is out of the 10-second audio segment.

\yz{The size of our token dictionary is 1265. Beyond task-specific query tokens and special tokens (\texttt{<sos>} for start of sentence, \texttt{<eos>} for end of sentence, \texttt{<pad>} for padding and \texttt{<unk>} for unknown), we encode time at 10ms resolution, yielding 1001 unique tokens for 10-second segments. Following~\cite{hawthorne2021sequence}, we use 128 tokens each for pitch and velocity. We also introduce a note-sustain" token, used when a note extends beyond the boundary of a training segment, and either its beginning or ending falls outside the current training area.}

\subsection{Dataset}
% \yz{We employ }the Maestro dataset~\cite{hawthorneenabling} contains around 200 hours of synchronized \yz{piano} and MIDI recordings from the past decade of the International Piano-e-Competition. The MIDI data was captured using high-precision Yamaha Disklaviers, and the recordings have been precisely aligned to within 3 milliseconds. The dataset is segmented into individual musical pieces, each annotated with the composer, title, and year of performance.

We employ the Maestro dataset~\cite{hawthorneenabling}, which contains about 200 hours of paired piano recordings and MIDI score, sourced from the International Piano-e-Competition. The score is directly captured from the Yamaha Disklaviers piano, and have been aligned to have an error margin of within 3 milliseconds against the recordings. The dataset is then segmented into individual musical pieces. We utilize the official train/validation/test split \yz{provided by the Maestro dataset}, \yz{where} the number of recordings and total duration in hours are 962/137/177 songs and 159.2/19.4/20.0 hours, respectively.

\subsection{Training and Evaluation Setup}
%% Yongyi: 09/03, dropping Hft for now in case of incomplete exps.

% 重新训练的实验设置
% We re-trained the CRNN and HPPNet by ourselves and used the original Hft-transformer checkpoint provided in its website. When retraining the CRNN and HPPNet, we set the frame length as 10 seconds and 20 seconds, respectively, and set the input audio length as 10 seconds.
\yz{The original CRNN model is trained on 10-second long audio segments, while the HPPNet model is trained on 20-second long ones. For a fair comparison, we retrain HPPNet on 10-second long segments and achieves comparable performance of reported in \cite{wei2022hppnet}. For the LM decoder, we use 6 transformer layers with 16 attention heads, and an embedding dimension of 1024. After the models are re-trained, we connect their output to the LM decoder through a linear layer, then train end-to-end using the AdamW optimizer with a learning rate of 1e-5, a batch size of 5, and a maximum step count of 1 million. All experiments are conducted on NVIDIA RTX 4090s. We train all settings until convergence is observed.}

\subsection{Metrics}
%rephrase 这两段
% In evaluating the performance of a piano transcription system, we use the Note $F_1$ score metric: the harmonic mean of precision and recall in detecting individual notes. This involves matching each predicted note with a unique ground truth note based on onset time, pitch, and optionally offset time. Additionally, onset velocity can be used to discard matches with drastically different velocities. We primarily use an $F_1$ score that takes into account onsets, offsets, and velocities. We also include results for $F_1$ scores that consider only onsets or onsets and offsets. We defer to the mir\_eval library for a precise definition of the (standard) transcription metrics we use.

We follow the default settings in \texttt{mir\_eval} to determine if a predicted note is correct. Compared to a ground-truth note, a correctly predicted note should have an onset within a window of $\pm 50$ milliseconds, a pitch within a window of $\pm 50$ cents, a velocity within a window of 0.1 after normalizing to the interval [0,1], and an offset within $\pm 50$ milliseconds or 20\% of the note's duration, whichever is larger.
% Offset prediction is more forgiving, as piano is a percussive instrument, and it is perceptually more important to accurately predict the onsets of notes rather than their offsets. 
Following previous literature~\cite{kong2021high, wei2022hppnet, hawthorne2021sequence}, we primarily use an $F_1$ score that takes into account `pitches, onsets, offsets, and velocities (On Off Vel $F_1$)'. We also include results for $F_1$ scores that consider `pitches and onsets (On $F_1$)' or `pitches, onsets and offsets (On Off $F_1$)'.
% we primarily use an $F_1$ score that takes into account onsets, offsets, and velocities. We also include results for $F_1$ scores that consider only onsets or onsets and offsets.}

% Because piano is a percussive instrument, it is generally easier (and also more perceptually important) to accurately identify note onsets compared to offsets. We use mir\_eval’s default match tolerance of 50 ms for onsets and the greater of 50 ms or 20\% of the note’s duration for offsets.

\section{Results and Discussions}
\label{sec:results}
% As shown in Tab. \ref{tab:result}, the "Hierarchy" version results surpass the "Roll" version in all of the three metrics, proving the effectiveness of LM decoder in piano transcription. Besides, the "Hierarchy" version results surpass the "Flatten" version shows that the hierarchical design of our model is effective. Moreover, by comparing the results among different encoders, we can conclude that our method can improve the results for nearly all kinds of the encoders that predict frame-level transcription results, and the better the encoder is, the better piano transcription results the LM Decoder will output.

\yz{Results are shown in Tab.~\ref{tab:result}. Comparing between the three settings, we can see that the flattened sequence is slightly worse at modeling onset behavior, but drastically worse at predicting note offset; whereas the hierarchical setting consistently reports comparable or better performance. It is worth noting that the ``Roll'' approach requires setting a threshold to gate notes as posterior information, whereas the ``Hierarchy'' approach do not. We have also achieved new state-of-the-art result of the LM-based piano transcription model with the HPPNet setting on all three $F_1$ scores.}

\begin{table}
  \caption{Experimental results. Best performing setting for each metric is shown in \textbf{bold}.}
  \label{tab:result}
  \centering
  \resizebox{\linewidth}{!}{%
    \begin{tabular}{l r ccc}
      \toprule
      Models & Params & On $F_1$ & On Off $F_1$ & On Off Vel $F_1$ \\
      \midrule
      Hawthorne et al.~\cite{hawthorne2021sequence} & 54M & 0.960 & 0.839 & 0.826 \\
      Hawthorne et al.~\cite{hawthorne2021sequence} & 213M & 0.956 & 0.828 & 0.814 \\
      \midrule
      CRNN Roll & 200M & \textbf{0.967} & 0.825 & 0.809 \\
      CRNN Flatten & 200M & 0.943 & 0.393 & 0.386 \\
      CRNN Hierarchy & 200M & 0.962 & \textbf{0.845} & \textbf{0.819} \\
      \midrule
      HPPNet Roll & 181M & 0.971 & 0.822 & 0.810 \\
      HPPNet Flatten & 181M & 0.950 & 0.390 & 0.382 \\
      HPPNet Hierarchy & 181M & \textbf{0.971} & \textbf{0.852} & \textbf{0.832} \\
      \bottomrule
    \end{tabular}%
  }
\end{table}

Comparing between our results and results from \cite{hawthorne2021sequence}, which also applies a flattened sequence during training, we see that their results are much closer to the ``Roll'' approach. We believe this is primarily due to the influence of sequence length. In~\cite{hawthorne2021sequence}, each segment is 4.088 seconds, while in ours, the segment length is 10 seconds. This resulted in more notes per sequence, and with the flattened approach, sequence length grows quickly with note events, which may have hindered model performance. Also, \cite{hawthorne2021sequence} used an encoder-decoder language modeling architecture, while we use a decoder-only language model. We hypothesize that the non-autoregressive nature of the encoder-decoder architecture helped it better encode information that would be especially lost in the long, flattened sequence~\cite{fu2023decoder}, therefore increased its performance compared to our flattened setting.

% We can learn from Tab. \ref{tab:compare_size} that the decoder size has little influence on the piano transcription results. So the piano transcription results are mainly influenced by how powerful the feature extraction capacity the encoder has.

% \subsection{Ablation Study}
%ablation study: flatten 的设置
% To prove the effectiveness of our model design, we firstly compare our results with that of all these three models (CRNN, HPPNet, and Hft-transformer) using rule-based post-processing to fuse frame-level onset, offset, pitch, and velocity output. Besides, we predict onset, pitch, velocity, and offset tokens in one long sequence as the "Flatten" version of our model. We compare our "Hierarchy" version model with the "Flatten" version to prove the superiority. \yz{For the ``Flatten'' version, we use separate dictionaries for onset and offset time.}

%不同大小decoder的设置
% We set the decoder size as XXXMB, XXXMB, XXXMB as small, medium, and large versions to compare the influence of the LM Decoder size to the piano prediction results. In detail, we set the transformer layer number as X,X,X, dimension size as X,X,X for the small, medium, and large versions, respectively.

% \subsection{Ablation Studies}

% \yz{We }

\begin{table}[bt!]
    \caption{Different LM Decoder Settings.}
    \centering
    \begin{tabular}{|l|c|c|c|c|c|}
        \toprule
        \textbf{Setting} & \textbf{\# Layer} & \textbf{\# Head} & \textbf{\# Embed Dim} & \textbf{Batch Size} \\
        \midrule
        Tiny & 4 & 8 & 512 & 64 \\
        Small & 6 & 12 & 768 & 56 \\
        Base & 6 & 16 & 1024 & 40 \\
        Large & 12 & 32 & 1024 & 24 \\
        \bottomrule
    \end{tabular}
    \label{tab:model-comparison}
\end{table}

\yz{Language models have been observed to show a ``scaling law'' where large models with more parameters exhibit better performance~\cite{kaplan2020scaling}. However, on the piano transcription task, \cite{hawthorne2021sequence} observed that larger model overfits rather than generalize better. To better examine this phenomenon, we conduct three scaling experiments by training hierarchical token sequence models on three settings, dubbed as ``tiny'', ``small'' and ``large'', as shown in Table~\ref{tab:model-comparison}. We conduct these scaling experiments on 8 NVIDIA RTX 4090s.}

\begin{table}
  \caption{Results for scaling experiments. Best results are shown in \textbf{bold}.}
  \label{tab:result_scaling}
  \centering
  
  \begin{tabular}{l r ccc}
    \toprule
    Models & Params & On $F_1$ & On Off $F_1$ & On Off Vel $F_1$ \\
    \midrule
    HPPNet tiny & 41M & 0.938 & 0.807 & 0.745 \\
    HPPNet small & 105M & 0.963 & 0.805 & 0.763 \\
    HPPNet base & 181M & \textbf{0.971} & \textbf{0.852} & \textbf{0.832} \\
    HPPNet large & 335M & 0.967 & 0.777 & 0.740 \\
    \bottomrule
  \end{tabular}
\end{table}

\begin{figure}[bt!]
    \centering
    \includegraphics[width=0.9\linewidth]{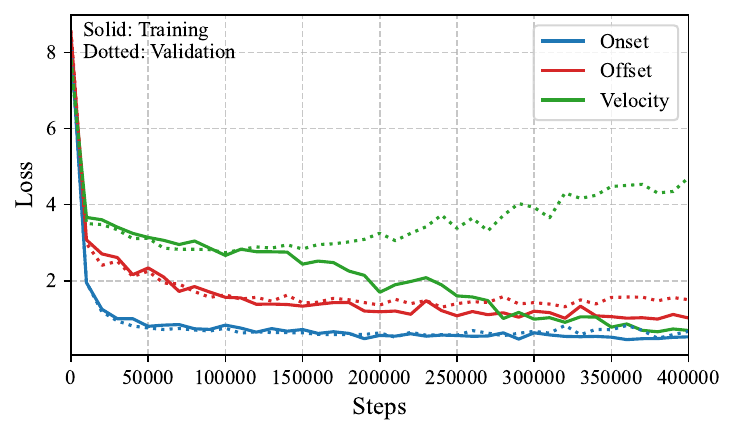}
    \caption{Onset-pitch, velocity and offset loss on the training set (solid lines) and validation set (dotted lines). Best viewed in color.}
    \label{fig:scaling_law}
\end{figure}

\yz{Results for the scaling experiments are shown in Table~\ref{tab:result_scaling}. We observe that compared to the original setting, no significant change in any metric is observed, while all settings slightly under-perform the ``base'' setting. To examine the training process, we plot the training and validations sets' loss at different timesteps in Figure~\ref{fig:scaling_law}. We observe that while onset-pitch and offset language models quickly plateaus during training, the velocity training show over-fitting at only about 100k steps, similar to the phenomenon described in~\cite{hawthorne2021sequence}. This suggests that the scaling behavior for different token types are different, and further merits our hierarchical approach.}

% \begin{table}
%   \caption{Result with different LM decoder size. Encoder e.g.: HPP (TBD)}
%   \vspace{6pt}
%   \label{tab:compare_size}
%   \centering
%   \begin{tabular}{lccc}
%     \toprule
%     & Onset & Offset & Vel \\
%     \midrule
%     Roll \\
%     Tiny \\
%     Small \\
%     Medium \\
%     Large \\
% \bottomrule
% \end{tabular}
% \end{table}

\section{Conclusions}
\label{sec:conclusions}
% \yz{In this paper, we propose a hirearchical language modeling approach that combines the strength of frame-level systems and language modeling approaches, which reduces sequence length and empirically shows improved performance. We present scaling analysis that shows the language model's size does not directly correlate with performance. We call for further investigation into improving the scalability of LM-based AMT systems.}

In this paper, we propose a hybrid method that combines pre-trained roll-based encoders with an LM decoder. Besides, our approach employs a hierarchical prediction strategy, first predicting onset and pitch, then velocity, and finally offset. The hierarchical prediction strategy reduces computational costs by breaking down long sequences. Evaluated on two benchmark roll-based encoders, our method outperforms traditional piano-roll outputs 0.01 and 0.022 in onset-offset-velocity F1 score, demonstrating its potential as a performance-enhancing plug-in for any roll-based encoder. Results show that encoder choice significantly impacts performance more than LM size, highlighting the importance of encoder selection in AMT. This study calls for further investigation into improving the scalability of LM-based AMT systems.

\section{Acknowledgement}
This research is supported by IdeaBooster funding from the Chinese University of Hong Kong with project code of IDBF24ENG07.

\bibliographystyle{IEEEtran}
\bibliography{references}

\end{document}